\documentclass{jetpl}
\usepackage{multirow}
\usepackage{supertabular}
\usepackage{epsfig}             
\usepackage{graphicx}
\usepackage[pdf]{pstricks}
\usepackage[T1]{fontenc}
\usepackage{longtable}
\usepackage{multirow}
\usepackage{gensymb}
\usepackage{cite}
\twocolumn

\lat

\title{Search for possible alpha-condensate states in $^{20}$Ne}

\rtitle{Search for possible alpha-condensate states in $^{20}$Ne}

\sodtitle{Search for possible alpha-condensate states in $^{20}$Ne}

\author{A.\,S.\,Demyanova$^{+}$\thanks{e-mail: a.s.demyanova@bk.ru},
A.\,N.\,Danilov$^+$, S.\,A.\,Goncharov$^{*}$, V.\,I.\,Starastsin$^{+\circ}$, T.\,I.\,Leonova$^{+\circ}$}

\rauthor{A.\,S.\,Demyanova, A.\,N.\,Danilov, ...}

\sodauthor{Demyanova, Danilov, ...}

\address{$^+$National Research Centre Kurchatov Institute,
123182 Moscow, Russia\\~\\
$^{*}$Lomonosov Moscow State University, 119991 Moscow, Russia\\~\\
$^{\circ}$National Research Nuclear University MEPhI (Moscow Engineering Physics Institute), 115409 Moscow, Russia\\~\\
}

\dates{*}{*}

\abstract{The root mean square radii of $^{20}$Ne in the short-lived excited states were experimentally deduced for the first time from the analyses of $\alpha$+$^{20}$Ne diffraction scattering. Differential cross sections of the elastic and inelastic $\alpha$+$^{20}$Ne scattering in the incident energy range from a few MeV/nucleon up to 100 MeV/nucleon were analyzed by the modified diffraction model. No significant radius enhancement for the members of K$^{\pi}$ = 0$_{1}^{+}$ and K$^{\pi}$ = 2$^{-}$ bands in comparison with the ground state was observed. At the same time 20\% radius enhancement was obtained for the K$^{\pi}$ = 0$_{1}^{-}$ band members. Moreover, for the 0$_{2}^{+}$ state located above $\alpha$-emission threshold increased radius was observed. This result can speak in favor of possible $\alpha$-condensate structure of the 0$_{2}^{+}$ state and can be considered as a possible analog of the famous 7.65-MeV 0$_{2}^{+}$ Hoyle state of $^{12}$C.}


\begin{document}

\maketitle

\section{Introduction}

An intensive study of four-nucleon correlations of the $\alpha$-cluster type initiated more than 50 years ago \cite{cite1} established their important role in nuclei. The microscopic $\alpha$-cluster models \cite{cite2, cite3, cite4, cite5, cite6} have succeeded in describing the structure of many states in light nuclei, in particular, around the threshold energy of breakup into constituent clusters. Considerable attention has been drawn to the studies of $\alpha$-cluster states in $^{12}$C, especially the second 0$^{+}$ state, located at E$_{\text{x}}$ = 7.65 MeV, which is 0.38 MeV above the 3$\alpha$ threshold. As early as 1954, Hoyle showed \cite{cite7} that this level plays an extremely important role in nucleosynthesis. The properties of the Hoyle state in $^{12}$C determine the ratio of carbon to oxygen formed in the stellar helium burning process that strongly affects the future evolution of stars. Detailed analyses of the structure of $^{12}$C with the microscopic 3$\alpha$ cluster model \cite{cite8, cite9} was made about 30 years ago. The 3$\alpha$ generator coordinate method (GCM) \cite{cite8} and 3$\alpha$ resonating group method (RGM) \cite{cite9} calculations showed that the 7.65 MeV 0$^{+}_{2}$ state in $^{12}$C has a loosely coupled 3$\alpha$ structure and an enlarged radius. Modern microscopic calculations in the framework of cluster models such as the antisymmetrized molecular dynamics (AMD) \cite{cite10} and the fermionic molecular dynamics (FMD) \cite{cite11} also predict an increased radius of this above-threshold cluster state. Much recent attention has been focused on experimental studies of the $\alpha$-cluster structure of the excited states in $^{12}$C as well as of the neighboring loosely bound $^{14}$C and $^{10}$Be nuclei \cite{cite12, cite13, cite14, cite15}.

A hypothesis of existence of $\alpha$-particle Bose-Einstein condensation ($\alpha$BEC) \cite{cite16} in finite nuclei which predicted appearance of nuclear states with unusual dilute $\alpha$-cluster structure resembling a gas of almost non-interacting $\alpha$-particles with considerably enhanced radii, has attracted plenty of attention in the two last decades. This idea gave impetus to further development of the theoretical approaches concerning $\alpha$-particle clustering initiated more than 50 years ago, and the experimental search for the diluted nuclear excited states. Testing the $\alpha$BEC model predictions became a challenge to experimental nuclear physics. In the context of the $\alpha$BEC hypothesis, the 7.65 MeV level in $^{12}$C \cite{cite16} is considered to be the simplest example of the $\alpha$-condensed state playing the role of a test for the whole problem. The direct measurement of the radius of this short-lived nuclear state is impossible, because of its very short half-life [$\tau_{1/2}$($0^{+}$, 7.65 MeV) $\sim$ 2×10$^{-16}$ s]. Estimates of the $\alpha$BEC model (R$_{\text{rms}}$(0$^{+}_{2}$) = 4.31 fm) \cite{cite17} is nearly twice as large as the radius of the ground state (2.34 fm). Nevertheless, the recent calculations in the framework of the ab initio Lattice Effective Field Theory (L-EFT) \cite{cite18} predicted the $rms$ radius (2.4 fm) almost equal to the $rms$ radius of $^{12}$C in the g.s. Application of our method, the Modified diffraction model (MDM) \cite{cite19} to the inelastic scattering of $^{2}$H, $^{3}$He, $^{4}$He, $^{6}$Li, and $^{12}$C on $^{12}$C in a wide range of energies allowed us to determine the consistent values of the $rms$ radii of $^{12}$C in the excited states up to E$_{\text{x}}$ = 11 MeV. In particular, we showed that in line with the expectations, the Hoyle state has an abnormally large radius, approximately 1.25 times larger than that for the ground state. The best agreement with this experimental result showed the AMD calculations \cite{cite20} R$_{\text{rms}}$(0$^{+}_{2}$) = 2.90 fm, and the no-core symplectic model (NCSpM) calculations \cite{cite21} 2.93 fm, while the $\alpha$BEC model result overestimates the radius of the Hoyle state

Another important characteristic predicted by the $\alpha$BEC model is a high probability W($\alpha$) (about 70–80\%) of the occupation of the lowest s orbit with zero angular momentum for all $\alpha$. We have estimated this value based on experimental data $\alpha$+$^{12}$C and obtained that in the Hoyle state, there exists a dominated (62\% of the occupation probability) configuration with all three $\alpha$ clusters on the s orbit \cite{cite22} Calculations of the single $\alpha$ orbital behavior in 0$^{+}$ states as a function of the nuclear radius R$_\text{N}$ carried out by Yamada and Schuck \cite{cite23} demonstrated a correlation between the values of the radius and $W(\alpha)$. $W(\alpha)$ = 70\% corresponds to a radius as large as 4.3 fm, whereas $W(\alpha)$ = 60\% suits to R$_\text{N}$ = 3 fm, the value obtained experimentally. These findings resolve apparent inconsistence between both the most important characteristics \big(the $rms$ radius and $W(\alpha)$\big). Taking into account all these results, we can conclude that the 0$^{+}_{2}$ Hoyle state possesses only some rudimentary features of $\alpha$-condensation if any.

A question naturally arises: do analogs of the Hoyle state exist in more massive 4$N$ nuclei. First possible candidate is the $^{16}$O. Really, the excited state of $^{16}$O located 570 keV above the 4$\alpha$-particle complete dissociation threshold (14.44 MeV) and 285 keV above the $\alpha$ + $^{12}$C$^*$(0$^{+}_2$) threshold (14.81 MeV), namely, the 15.1-MeV 0$^{+}_6$ state, which was discovered by Marvin and Singh \cite{cite24} as early as 1972, now is considered as the most probable candidate to be an analog of the Hoyle state. Suggestions about the structure of the 15.1-MeV 0$^{+}_6$ state were proposed also in the framework of $\alpha$BEC [16]. Funaki et al. realized \cite{cite25, cite26} four-body orthogonality condition model (OCM) calculations of the $^{16}$O energy spectrum using the $\alpha$-particle condensation wave functions and found that the 0$^{+}_6$ state has the “gigantic” $rms$ radius R$_\text{rms}$= 5.6 fm comparable to the radius of the uranium nucleus. For the 13.6-MeV 0$^{+}_4$ state first found by Wakasa et al. \cite{cite27}, the calculations \cite{cite25, cite26} also showed surprising enhancement of radius. However, no experimental information about the size of $^{16}$O in these states was obtained. In our work \cite{cite28}, inspired by the appearance of recent experimental data on $\alpha$ + $^{16}$O inelastic scattering at 386 MeV \cite{cite29}, we determined the radii of $^{16}$O in the 0$^{+}_4$, 0$^{+}_5$, and 0$^{+}_6$ states located near and above the 4$\alpha$-particle dissociation threshold deduced from the analysis of these data by means of the MDM. No significant radius enhancement in any states was observed. This result concerns, first, the 0$^{+}$ states located in the vicinity of 4$\alpha$-particle dissociation threshold. In particular, we did not confirm the existence of a dilute state with super-large radius associated with the 15.1-MeV 0$^{+}_6$ state, which was predicted by the $\alpha$BEC model. The $rms$ radius of $^{16}$O in this state was found similar to the radius of $^{16}$O in the ground state. From this point of view, the 0$^{+}_6$ state cannot be considered as an analog of the Hoyle state in $^{12}$C.

The next candidate is $^{20}$Ne. It was widely studied both theoretically and experimentally in literature. Below is the short overview of existing works. 
In \cite{cite30} a wide range of 4$N$ alpha cluster nuclei were studied, among them was $^{20}$Ne. Predicted structure of the ground and excited states of $^{20}$Ne are shown in Fig. 1.

\begin{figure}
	\centering	
	\includegraphics[width=.85\columnwidth]{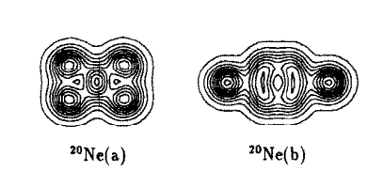}
	\caption{Fig. 1. Cluster density contours from\cite{cite30}: a – g.s., b – excited states.}
	\label{Fig:1}
\end{figure}


R. Bijker and F. Iachello \cite{cite32} studied the cluster structure of $^{20}$Ne and proposed that the available experimental data can be well described by a bi-pyramidal structure with $\mathcal{D}_{3h}$ symmetry. According to calculations \cite{cite32}, low-lying excited states can be distributed among vibrationless ground state band, nine vibrational modes expected on the basis of $\mathcal{D}_{3h}$ symmetry (3 singly degenerate and 3 doubly degenerate), and of six of the double vibrational bands expected on the basis of $\mathcal{D}_{3h}$ symmetry. According to the authors \cite{cite32}, strong evidence for the occurrence of this symmetry comes from the observation all observed K$^\pi$ bands with band-heads up to an excitation energy of about 12 MeV.

Y. Kanada-Enyo and K. Ogata within AMD \cite{cite33} investigated the structure and transition properties of the K$^\pi$ = 0$^{+}_1$, K$^\pi$ = 2$^-$, and K$^\pi$ = 0$^{-}_1$ bands of $^{20}$Ne via proton and $\alpha$ scattering on $^{20}$Ne. These 3 bands correspond to bands predicted in \cite{cite32}. In the structure calculation of $^{20}$Ne with AMD, $^{16}$O + $\alpha$ cluster structures were obtained in the parity-doublet K$^\pi$ = 0$^{+}_1$ and K$^\pi$ = 0$^{-}_1$ bands. While for the K$^\pi$ = 2$^{-}$ band, the $^{12}$C + 2$\alpha$ -- like structure was obtained. The $rms$ radii were determined for all band members: for the K$^\pi$ = 0$^{+}_1$ band -- normal non increased radii, for the K$^\pi$ = 0$^{-}_1$ band -- increased by $\sim$ 0.2 fm radii, for the K$^\pi$ = 2$^{-}$ band -- normal non increased radii.

S. Adachi and collaborators conducted \cite{cite34} the coincidence measurement of $\alpha$ particles inelastically scattered from $^{20}$Ne at 0$\degree$ and decay charged particles in order to search for the alpha-particle condensed state. It was found that the newly observed states at E$_\text{x}$ = 23.6, 21.8, and 21.2 MeV in $^{20}$Ne are strongly coupled to a candidate for the 4$\alpha$ condensed state in $^{16}$O. By author’s opinion, this result presents the first strong evidence that these states are the candidates for the 5$\alpha$ condensed state. However, spins and parities of these states are still ambiguous. 

Eventually, $^{20}$Ne was studied by ab initio calculations in \cite{cite35}. The formalism was demonstrated in a study of the $^{16}$O + $\alpha$ cluster structure of $^{20}$Ne, through inspection of the relative motion wave functions of the clusters within the $^{20}$Ne ground state and the 1$^-_1$ resonance. For the first time within a no-core shell model framework, the $\alpha$ partial width for the 1.06-MeV 1$^-$ resonance of $\alpha$ + $^{16}$O from ab initio SA-NCSM calculations of the $^{20}$Ne states were determined and a good agreement with experiment was shown. The importance of correlations in developing cluster structures was shown.

Approach proposed by our group is based on measuring the nuclear radii. In our group, several methods are developed to be used for measuring radii of nuclei in the short-lived excited states: the MDM, the Asymptotic normalization coefficients method \cite{cite36}, and the nuclear rainbow method (NRM) \cite{cite37}. In current article results of MDM application to existing literature data on $\alpha$ + $^{20}$Ne scattering is present.

\section{Results and discussion}

In order to measure nuclear radii in unbound states, we propose to use the MDM \cite{cite19} for the analysis of inelastic differential cross sections.

We found a large amount of inelastic scattering data $\alpha$ + $^{20}$Ne in a wide energy range 30-400 MeV \cite{cite34, cite38, cite39, cite40}. However, we haven’t found angular distributions for the states above 8 MeV excitations in literature. So hypothesis proposed in \cite{cite34} we can’t check via MDM. Angular distributions are present for the g.s. and low-lying excited states 1.63 MeV, 4.25 MeV, 4.97 MeV, 5.62 MeV, 5.79 MeV, 6.73 MeV, 7.17 MeV. But not all states can be analyzed within MDM. MDM currently can’t be applied to angular distributions with $L$ > 3, so 4.25 MeV ($L$ = 4) can’t be analyzed within MDM. The 4.97 MeV state has unusual spin-parity 2$^-$, so it can be excited only by two step process. By this reason, MDM can’t be applied to this state.

MDM is described in details in \cite{cite19}. MDM application to data on elastic and inelastic scattering made it possible to obtain consistent values for the root-mean-square radii of a large number of excited states via experimentally determining the diffraction radii for excited and ground states.

Let’s briefly repeat main aspects of the MDM. The idea of the MDM consists in assuming that the difference of the root mean-square radii of an excited and the ground state is equal to the difference of the respective diffraction radii:

\begin{equation}
    R^{*}_\text{rms} = R^{0}_\text{rms} + \big[R^{*}_\text{dif} - R_\text{dif}(0)\big],
\end{equation}

where $R^{0}_\text{rms}$ is the root-mean-square radius of the nucleus under study in the g.s. (it is assumed to be known) and $R^{*}_\text{dif}$ and $R_\text{dif}(0)$ are the diffraction radii calculated on the basis of the positions of the minima and maxima in the experimental angular distributions for, respectively, inelastic and elastic scattering. For the application of the modified diffraction model to be successful, it is necessary to determine reliably the angular-momentum transfer in inelastic scattering and the positions of diffraction extrema. In order to reach a satisfactory precision in determining diffraction radii, it is usually sufficient to observe two or three pairs of the minima and maxima. We also note that, for the application of the modified diffraction model to be legitimate, fulfillment of the so-called adiabatic condition is necessary. This means that the ratio of the initial energy in the c.m. frame, $E_\text{c.m.}$, to the energy of the state being considered, $E^*$, should satisfy the condition $E_\text{c.m.}$/$E^*$ $\gg$ 1 (in practice, a value of four to five is usually sufficient).

So, starting point is the diffraction radius $R_\text{dif}(0)$ from elastic scattering. Experimental elastic scattering data is present in Fig. 2. For convenience graphs are built as a function of linear transferred momentum. Diffraction radius is inversely proportional to linear transferred momentum $q$. As can be seen from Fig. 2, positions of minima/maxima are slowly systematically shifting to smaller $q$ value with energy decrease. This fact corresponds to diffraction radius dependence on energy: diffraction radius is decreasing with energy increase.  

\begin{figure}
	\centering	
	\includegraphics[width=1\columnwidth]{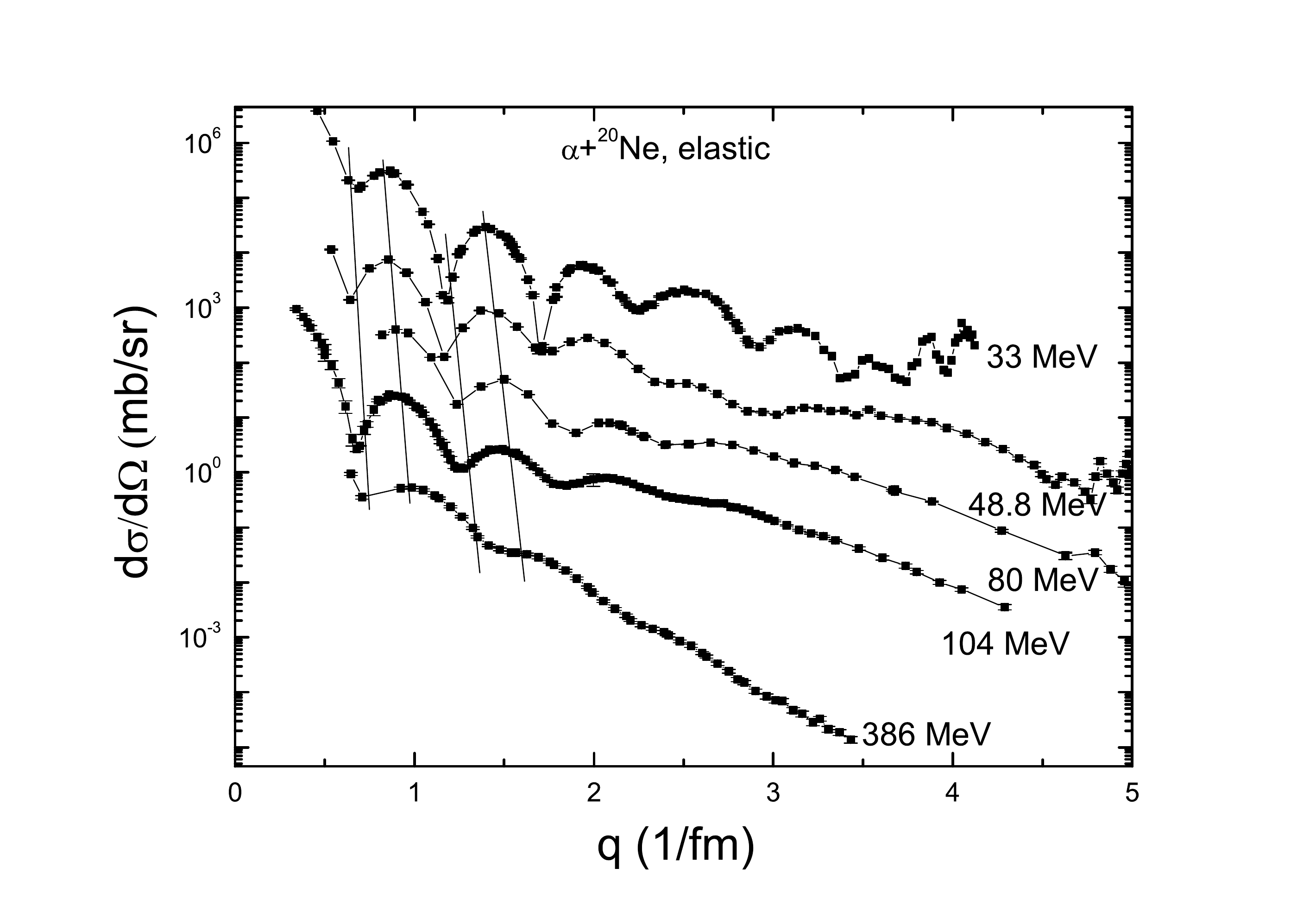}
	\caption{Fig. 2. Differential cross sections for elastic $\alpha$ + $^{20}$Ne scattering at $E(\alpha)$ = 33 and 80 MeV from \cite{cite38}, 48.8 MeV from \cite{cite39}, 104 MeV from \cite{cite40}, 386 MeV from \cite{cite34}.}
	\label{Fig:2}
\end{figure}

The diffraction radius of the ground state decreases linearly with energy increase from $R_\text{dif}(0)$ = 6.00 $\pm$ 0.02 fm at $E(\alpha)$ = 33 MeV to $R_\text{dif}(0)$ = 5.0 $\pm$ 0.1  fm at $E(\alpha)$ = 386 MeV.
Value of the g.s. radius $R_\text{rms}^0$ = 2.87 $\pm$ 0.03 fm is taken from \cite{cite41}.

\subsection{Band K$^\pi$ = 0$^{+}_1$}

The K$^\pi$ = 0$^{+}_1$ band contains states g.s. 0$^+$ -- 1.63 MeV 2$^+$ -- 4.25 MeV 4$^+$. As mentioned above, within MDM radius can be calculated for the 1.63 MeV 2$^+$ state. 

Experimental angular distributions for this state are present in Fig. 3.

\begin{figure}
	\centering	
	\includegraphics[width=1\columnwidth]{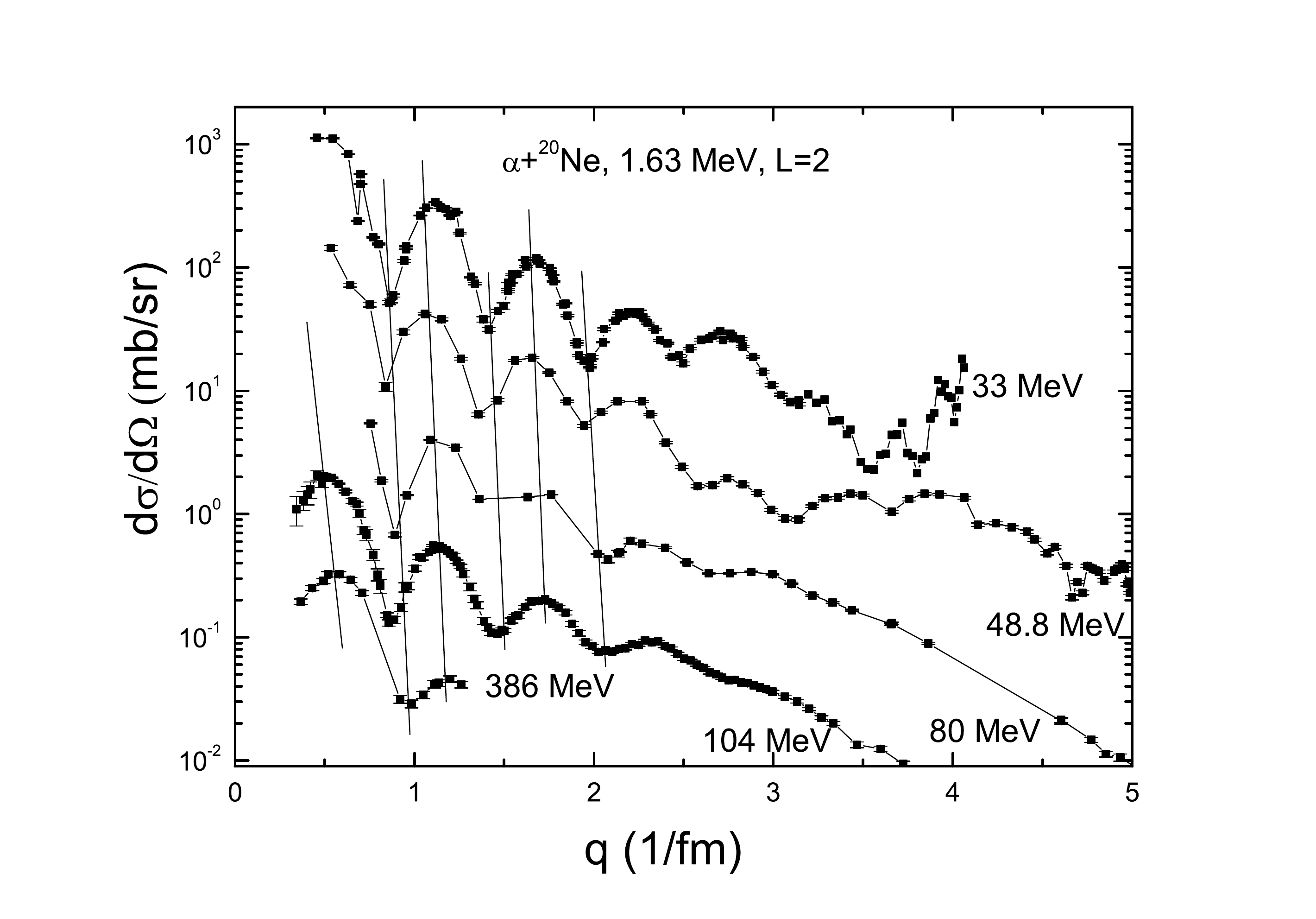}
	\caption{Fig. 3. Differential cross sections for inelastic $\alpha$ + $^{20}$Ne scattering with excitation of the 1.63 MeV state at $E(\alpha)$ = 33 and 80 MeV from \cite{cite38}, 48.8 MeV from \cite{cite39}, 104 MeV from \cite{cite40}, 386 MeV from \cite{cite34}.}
	\label{Fig:3}
\end{figure}

Averaged value of $rms$ radius obtained within MDM is $R_\text{rms}$(1.63) = 3.0 $\pm$ 0.2 fm. Within errors this value coincides with radius of the g.s. and value obtained within AMD analysis 3.01 fm. Members of this band have normal non-increased radii and by AMD predictions have $^{16}$O + $\alpha$ cluster structure.

\subsection{Band K$^\pi$ = 2$^{-}$}

The K$^\pi$ = 2$^{-}$ band contains states 4.97 MeV 2$^{-}$ -- 5.62 MeV 3$^{-}$ -- 7.00 MeV 4$^{-}$ -- 8.45 MeV 5$^{-}$. Due to MDM peculiarities, radius can be obtained only for the 5.62 MeV 3$^{-}$ state. Angular distributions for this state are present at 2 energies 33 (see in Fig. 4) and 50 MeV. Averaged value obtained using MDM is $R_\text{rms}$(5.62) = 2.9 $\pm$ 0.1 fm. Within errors this value coincides with radius of the g.s. and value obtained within AMD analysis 2.96 fm. Members of this band have normal non-increased radii and by AMD predictions have $^{12}$C + 2$\alpha$ cluster structure \cite{cite33}.

\begin{figure}
	\centering	
	\includegraphics[width=1\columnwidth]{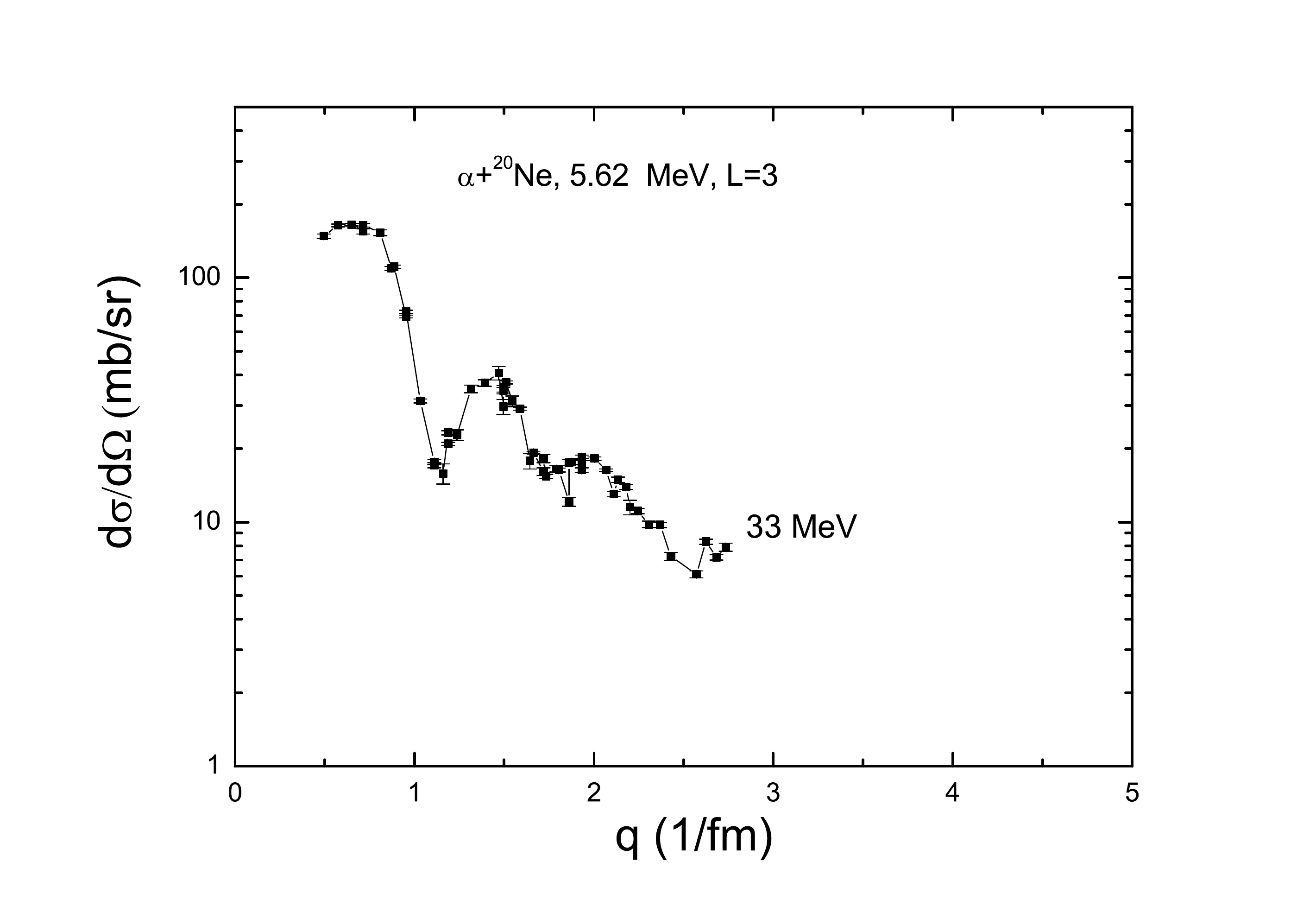}
	\caption{Fig. 4. Differential cross sections for inelastic $\alpha$ + $^{20}$Ne scattering with excitation of the 5.62 MeV state at $E(\alpha)$ = 33 from \cite{cite38}.}
	\label{Fig:4}
\end{figure}

\subsection{Band K$^\pi$ = 0$^{-}_{1}$}

The K$^\pi$ = 0$^{-}_{1}$ band contains states 5.79 MeV 1$^-$ -- 7.17 MeV 3$^-$ -- 10.26 MeV 5$^-$. Within MDM radius can be obtained for the first two excited states. Experimental angular distributions for these states are present in Fig. 5 and Fig. 6.

In \cite{cite40} 5.62 and 5.79 MeV states were not divided. Our DWBA analysis showed that the 5.79 MeV state dominates.  

\begin{figure}
\includegraphics[width=1\columnwidth]{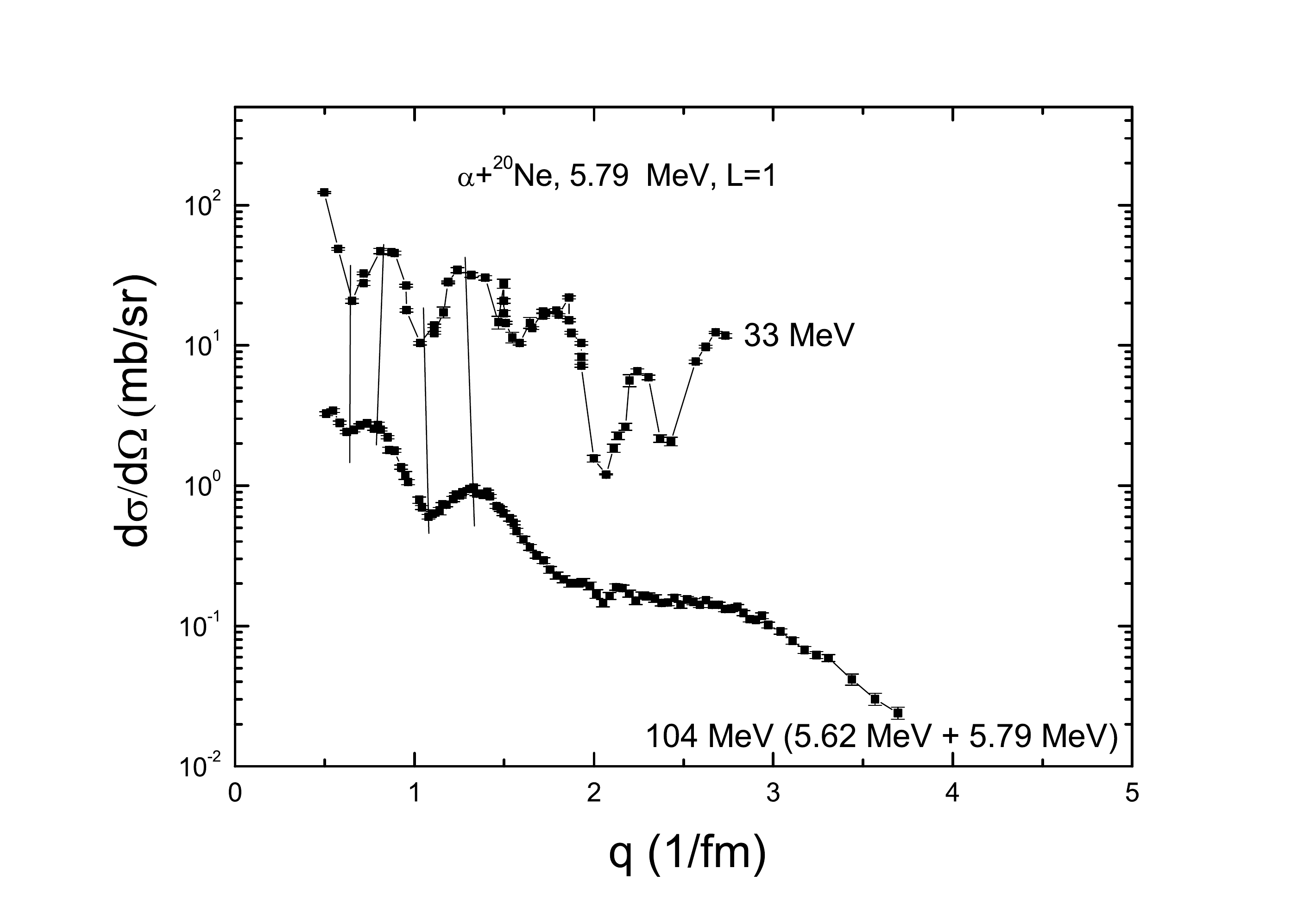}
\caption{Fig. 5. Differential cross sections for inelastic $\alpha$ + $^{20}$Ne scattering with excitation of the 5.79 MeV state at $E(\alpha)$ = 33 from \cite{cite38} and sum of states 5.62 and 5.79 MeV at $E(\alpha)$ = 104 MeV from \cite{cite40}.}
\label{Fig:5}
\end{figure}

\begin{figure}
\includegraphics[width=1\columnwidth]{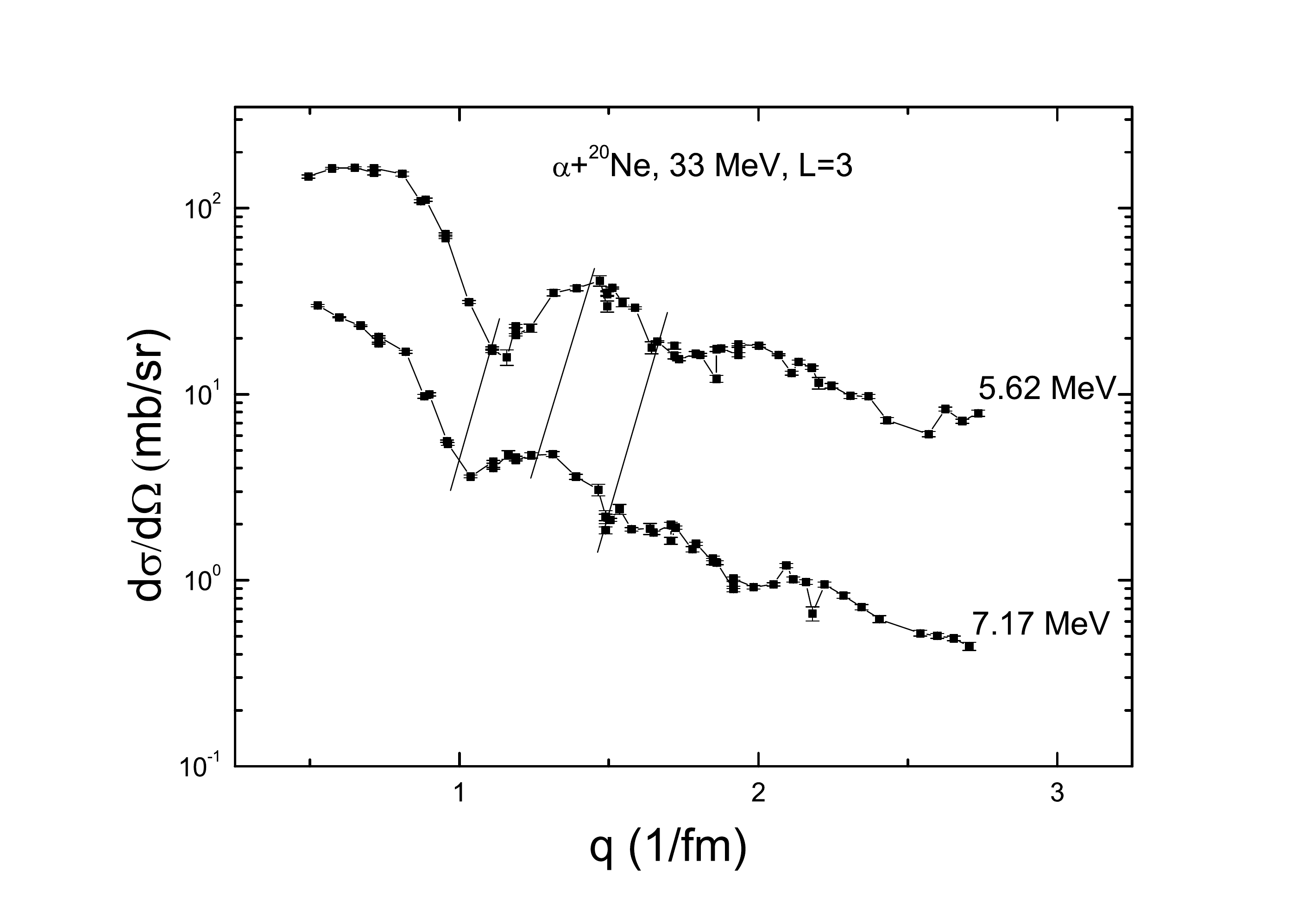}
\caption{Fig. 6. Comparison of differential cross sections for inelastic $\alpha$ + $^{20}$Ne scattering with excitation of the 5.62 and 7.17 MeV states at $E(\alpha)$ = 33 from \cite{cite38}.}
\label{Fig:6}
\end{figure}

Averaged radii obtained using MDM for both states are $R_\text{rms}$(5.79) = 3.4 $\pm$ 0.2 fm and $R_\text{rms}$(7.17) = 3.4 $\pm$ 0.2 fm. These values are increased comparing to radius of the g.s. and within errors coincide with values predicted by AMD 3.2 and 3.19 fm correspondingly. Moreover, as can be seen in Fig. 6, if we compare angular distributions for the 5.62 and 7.17 MeV states (both states are excited by transferred momentum $L$ = 3), then minima and maxima for the 7.17 MeV state are shifted to smaller $q$ values. This means that the 7.17 MeV state has increased radius. By AMD predictions this band has $^{16}$O + $\alpha$ cluster structure.

\subsection{Band K$^\pi$ = 0$^{+}_{2}$}

According  \cite{cite32}, the K$^\pi$ = 0$^{+}_{2}$ band contains states 6.73 MeV 0$^{+}$ -- 7.42 MeV 2$^{+}$ -- 9.03 MeV 4$^{+}$. We found angular distributions only for the 6.73 MeV state at single energy \cite{cite34}. Angular distribution contains only 4 points, so we made only estimate of the radius $R_\text{rms}$(6.73) $\sim$ 3.6 fm. 

\begin{figure}
	\centering	
	\includegraphics[width=1\columnwidth]{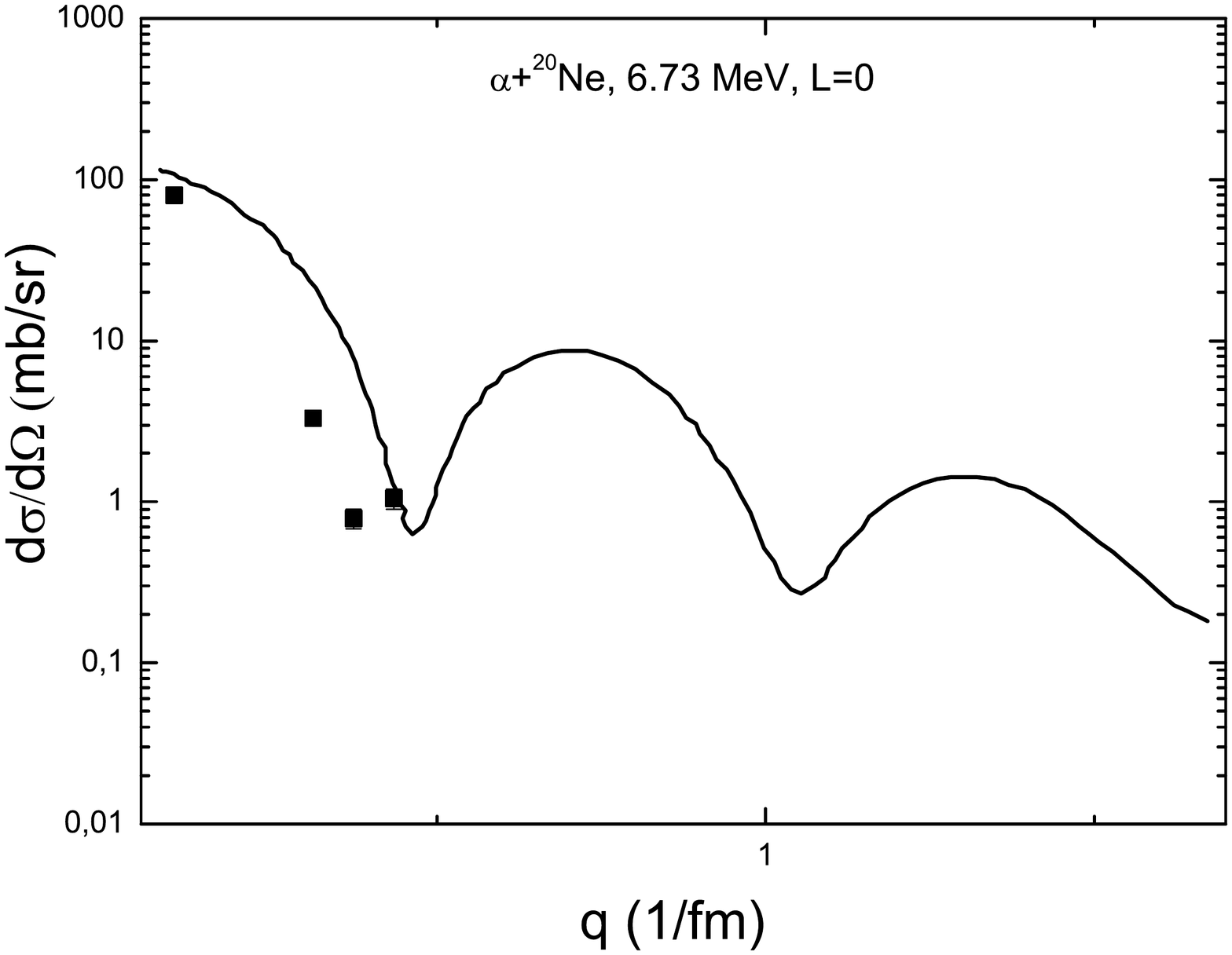}
	\caption{Fig. 7. Differential cross sections for inelastic $\alpha$ + $^{20}$Ne scattering with excitation of the 6.73 MeV at $E(\alpha)$ = 386 MeV from \cite{cite34} together with theoretical calculations from \cite{cite34}. }
	\label{Fig:7}
\end{figure}

We obtained 25\% radius increase for this state comparing to the g.s. radius. Practically the same increase we observed for the Hoyle state in $^{12}$C. The 6.73 MeV state is located 2 MeV above the $\alpha$-emission threshold, so our result can be the argument for the possible alpha-condensate structure. But further analysis is needed. New experimental data are required. Also it will be good to check next member of this band, the 7.42 MeV 2$^+$ state, and next 0$^+$ states (0$^{+}_{3}$, 0$^{+}_{4}$, etc) but there are no angular distributions in literature for them. 

Discussed in article rotational bands of $^{20}$Ne are present in Fig. 8 together with obtained within MDM radii. 
\begin{figure}
	\centering	
	\includegraphics[width=1\columnwidth]{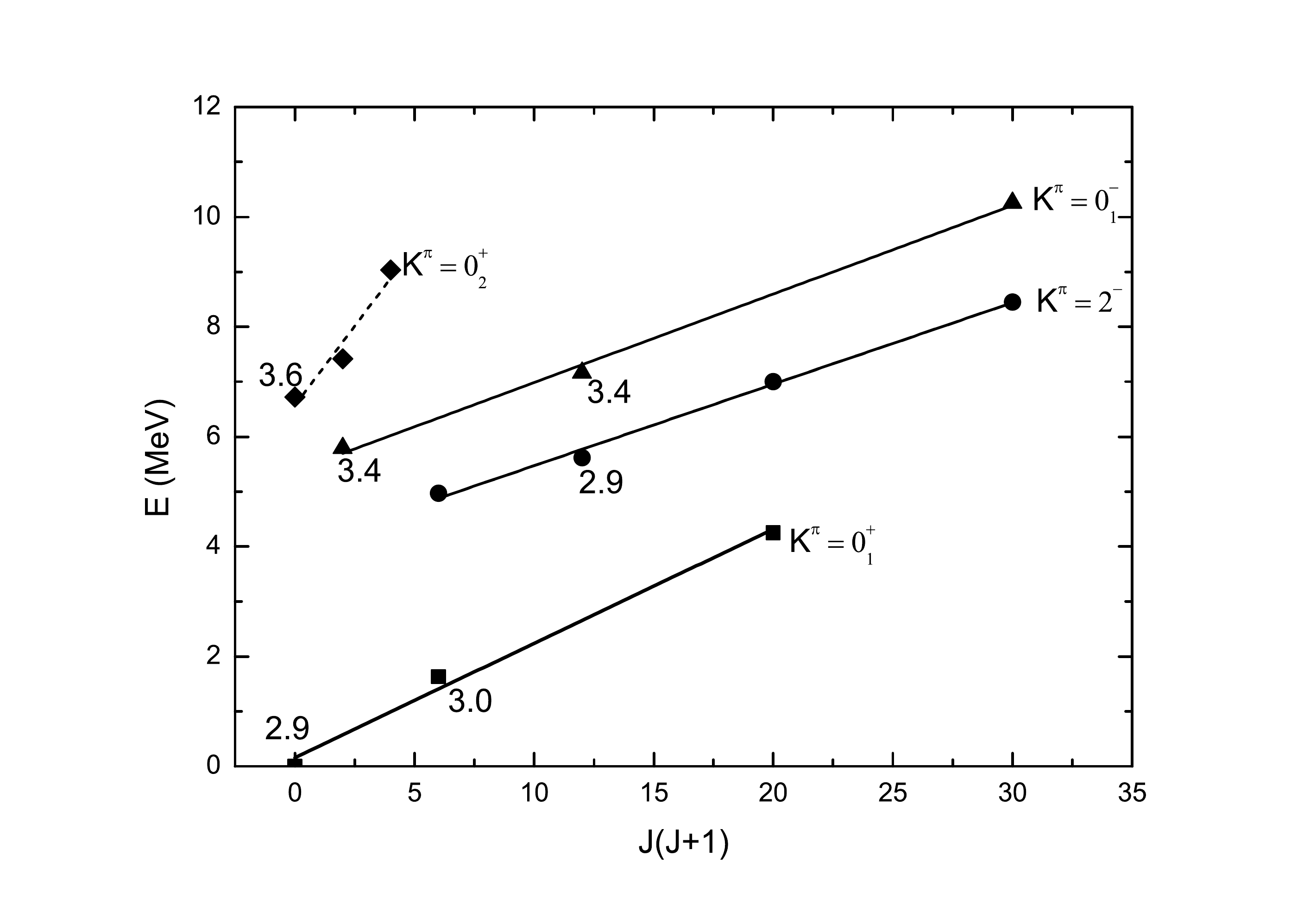}
	\caption{Fig. 8. First rotational bands of $^{20}$Ne: K$^\pi$ = 0$^{+}_1$ (squares), K$^\pi$ =0$^{-}_1$ (triangles), K$^\pi$ =2$^{-}$ (circles), K$^\pi$ = 0$^{+}_2$  (rhombuses)}
	\label{Fig:8}
\end{figure}

\section{Conclusions}

To conclude, the literature differential cross sections of the inelastic $\alpha$ + $^{20}$Ne scattering in the energy interval from a few tens MeV up to 400 MeV were analyzed. We determined directly the $rms$ radii of $^{20}$Ne in a number of states with excitation energies up to 7 MeV applying the MDM. No significant radius enhancement was observed for the members of K$^\pi$ = 0$^{+}_{1}$ and K$^\pi$ = 2$^{-}$ bands in comparison with the ground state. By AMD predictions, these bands have different structures: $^{16}$O+$\alpha$ for the K$^\pi$ = 0$^{+}_{1}$ band and $^{12}$C+2$\alpha$ for the K$^\pi$ = 2$^{-}$ band. At the same time 20\% radius enhancement was obtained for the K$^\pi$ = 0$^{-}_{1}$ band members. By AMD predictions, these band members have $^{16}$O+$\alpha$ structure. Moreover, our estimates of the radius of the $0_{2}^+$ state, the head of the K$^\pi$ = 0$^{+}_{2}$ band, showed 25\% radius increase. This state is located above $\alpha$-emission threshold. Obtained result can speak in favor of possible $\alpha$-condensate structure of the 0$^{+}_{2}$ state and can be considered as a possible analog of the famous 7.65-MeV 0$^{+}_{2}$ Hoyle state of $^{12}$C. 
Finally we hope that the results reported in this paper will lead to more understanding of the problem of nuclear radii in the excited states and will stimulate further experimental and theoretical studies of the properties of light nuclei.

\section{Acknowledgements}
The reported study was funded by the NRC Kurchatov Institute (№2767 from 28.10.21).


\end{document}